\documentclass[twocolumn,showpacs,preprintnumbers,amsmath,amssymb]{revtex4}
\input epsf
\usepackage{graphicx}% Include figure files
\usepackage{dcolumn}% Align table columns on decimal point
\usepackage{bm}% bold math

\newcommand{\be}{\begin{equation}}
\newcommand{\ee}{\end{equation}}
\newcommand{\bea}{\begin{eqnarray}}
\newcommand{\eea}{\end{eqnarray}}

\begin{document}

\title{Phase-control of directed diffusion in a symmetric optical lattice}

\author{M. Schiavoni, L. Sanchez-Palencia, F. Renzoni and G. Grynberg \\}
\affiliation{Laboratoire Kastler-Brossel, D\'epartement de Physique de l'Ecole
Normale Sup\'erieure, 24, rue Lhomond, 75231, Paris Cedex 05,
France.}

\date{\today}

\begin{abstract}
We demonstrate the phenomenon of directed diffusion in a symmetric periodic
potential. This has been realized with cold atoms in a one-dimensional
dissipative optical lattice. The stochastic process of optical pumping leads
to a diffusive dynamics of the atoms through the periodic structure, while a
zero-mean force which breaks the temporal symmetry of the system is applied
by phase-modulating one of the lattice beams. The atoms are set into
directed motion as a result of the breaking of the temporal symmetry of
the system.
\end{abstract}


\maketitle

It is now about two centuries that scientists observe and model the motion
of microscopic particles in a fluctuating environment. The year 1828 can
probably be indicated as the birth date of this field of research, with the
observation by Brown \cite{brown} of the random motion of particles in a fluid.
And it took about a century before that this phenomenon, now known as Brownian
motion, was modeled by Einstein \cite{albert}. More recently, the problem of
modeling molecular motors \cite{prost97}, i.e. microscopic objects moving
unidirectionally along periodic structures, has renewed the interest in the
field and stimulated much theoretical work devoted to the study of the
directed motion in a fluctuating environment in the absence of bias
forces. Molecular motors have been modeled by an asymmetric potential
(ratchet) and non-gaussian noise \cite{magnasco}. Unidirectional motion in
a ratchet potential is also obtained with gaussian noise and an applied
periodic force of zero average \cite{magnasco,prost94,hanggi,reimann}.

In this work we demonstrate the phenomenon of directed diffusion (DD), i.e.
directed motion in a fluctuating environment, in a {\it symmetric}
optical lattice. Consider the diffusive dynamics in a periodic potential
$U(x)$ of period $\lambda$, $U(x+\lambda)=U(x)$, in the presence of a
driving force $F(t)$ of period $T$, $F(t+T)=F(t)$. If the system is
symmetric in the sense that $U(-x)=U(x)$ and $F(t+T/2)=-F(t)$, there is no
net average transport through the periodic structure
\cite{prost94,rabitz,flach,flach2}.
Therefore to observe directed motion the spatiotemporal symmetry of the system
has to be broken. For a spatially symmetric potential, the symmetry of the
system can be broken by applying a non-monochromatic driving force containing
both odd and even harmonics. In the present investigation the driving force
has two components of frequencies $\omega$ and $2\omega$ and phase difference
$\phi$. We will demonstrate experimentally the phenomenon of DD for such a
configuration, with the phase $\phi$ playing the role of control parameter
for the amplitude and sign of the current of atoms through the lattice.

\begin{figure}[ht]
\begin{center}
\mbox{\epsfxsize 3.in \epsfbox{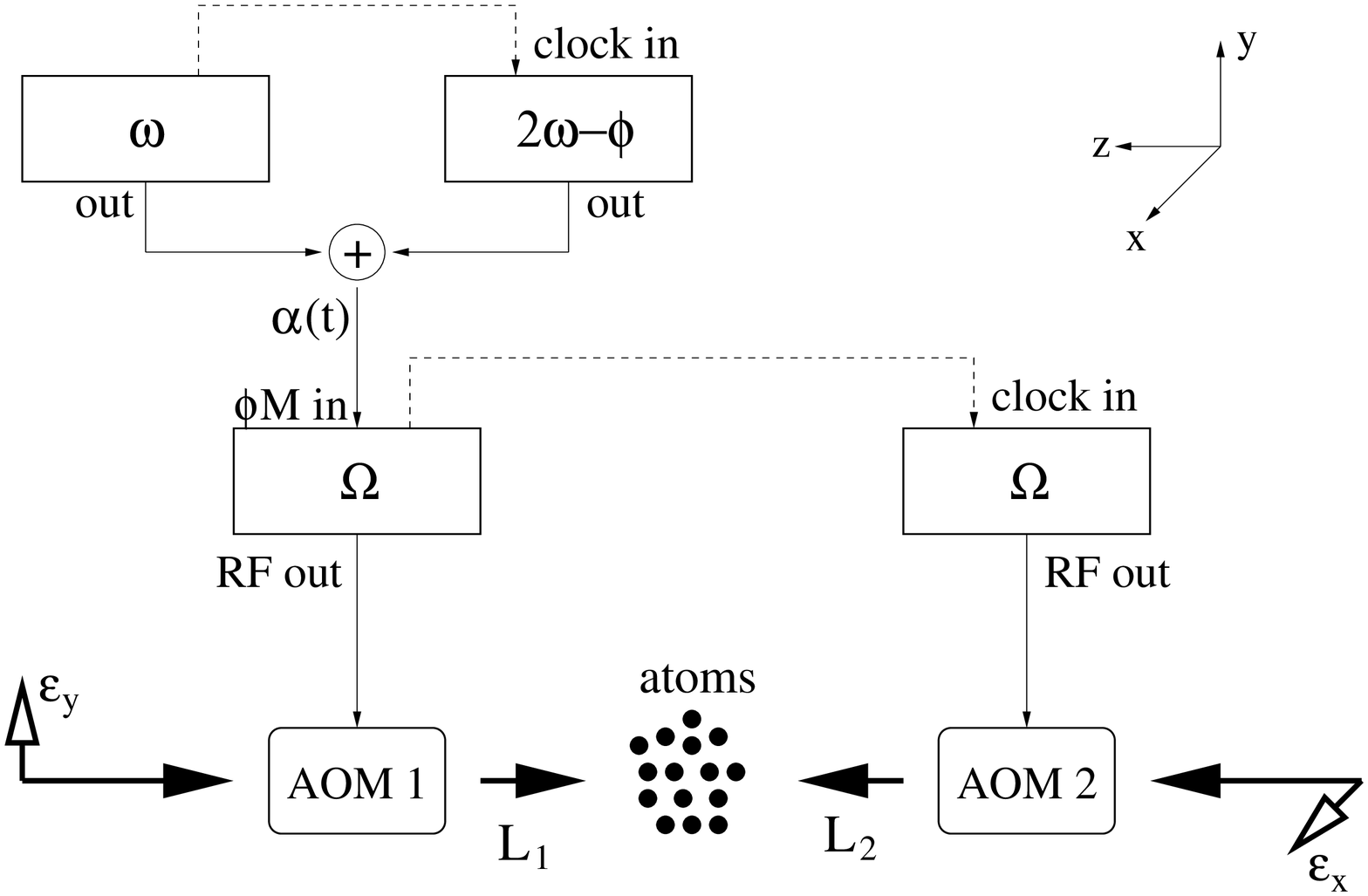}}
\end{center}
\caption{Sketch of the experimental setup. The phase of the laser field
$L1$ is $-kz-\omega_L t+\alpha (t)$, while the phase of the field $L2$
is $kz-\omega_L t$. Here $\omega_L$ includes the frequency shift $\Omega$
produced by the acousto-optical modulators. }
\label{fig1}
\end{figure}

Our symmetric periodic potential corresponds to a one-dimensional
lin$\perp$lin optical lattice \cite{robi}. The periodic structure is
determined by the interference of two counterpropagating laser beams
(L$_1$ and L$_2$), with crossed linear polarizations
(Fig. \ref{fig1}). This arrangement results in a periodic modulation of
the light polarization, which produces a periodic modulation of the light
shifts of the different ground states of the atoms. In this way an atom
experiences a periodic potential ({\it optical potential}), whose
amplitude and phase depend on the internal state of the atom. This
dependence allows Sisyphus cooling \cite{robi} to take place. Indeed,
the optical pumping between the different atomic ground states combined
with the spatial modulation of the optical potential leads to the cooling
of the atoms and to their localization at the minima of the optical
potentials, thus producing a periodic array of trapped atoms. The transport
of atoms through the lattice is determined by the optical pumping between
different ground state sublevels.
In fact, atoms at the bottom of a potential well strongly
interact with the light and therefore undergo fluorescence cycles. The
stochastic process of optical pumping may transfer an atom from a
potential well to a neighbouring one corresponding to a different optical
potential. This results in the transport of atoms through the lattice.
More precisely, in a wide range of lattice parameters the atomic dynamics
corresponds to normal diffusion \cite{epjd1,epjd2}.

In order to generate a time-dependent homogeneous force, we apply a
phase-modulation to one of the lattice beams, so that to obtain the
electric field configuration
\bea
& & \vec{E}=E_0 {\rm Re} \Big\{ \vec{\epsilon}_x \exp{[i(kz-\omega_L t)]} \\
& & \phantom{aaaaaaii} + \vec{\epsilon}_y \exp{[i(-kz-\omega_L t+\alpha (t))]}  \Big\} ~. \nonumber
\eea
Here $E_0$ is the (real) amplitude of the electric field, $k$ and $\omega_L$
the lattice-field wavevector and frequency, respectively. The modulated phase
is $\alpha (t)$. In the laboratory reference frame this laser configuration
generates a moving optical potential $U(2kz-\alpha (t))$. To be explicit,
consider the case of a $J_g=1/2\to J_e=3/2$ transition, which is the simplest
atomic transition for which Sisyphus cooling takes place. In this case the
moving bi-potential for the $|g,m=\pm 1/2\rangle$ ground states is
$U_{\pm}(2kz-\alpha (t))$ with $U_{\pm}(\xi)=U_0[-2\pm\cos\xi ]$, $U_0$
being the depth of the potential wells. Consider now the dynamics in the
moving reference frame defined by the coordinate transformation
$z'=z-\alpha (t)/2k$.  In this accelerated reference frame the optical
potential is stationary. In addition to this potential, the atom, of mass
$M$, experiences also an inertial force $F$ in the $z$-direction
proportional to the acceleration $a$ of the moving frame \cite{landau,maxim}:
\begin{equation}
F=-Ma=-\frac{M}{2k}\ddot{\alpha} (t)~.
\end{equation}
By choosing a phase modulation of the form
\begin{equation}
\alpha (t) = \alpha_0 [ A\cos (\omega t) +\frac{B}{4}\cos (2\omega t-\phi )]
\label{alpha}
\end{equation}
with $\phi$ constant, we obtain the inertial force
\begin{equation}
F=\frac{M\omega^2\alpha_0}{2k}\left[ A\cos (\omega t)+
B\cos (2\omega t-\phi )\right]
\label{force}
\end{equation}
which is the sum of two forces oscillating at the frequencies $\omega$ and
$2\omega$, with phase difference $\phi$. Hence, in the accelerated frame the
atoms cooled and trapped in the optical lattice experience a force containing
both even and odd harmonics, so that our system is suitable for the
observation of DD. All the results presented in this work are obtained in the
regime of non-adiabatic driving, with the frequency $\omega$ of the driving
force about equal to the frequency $\Omega_v$ of oscillation of the atoms at
the bottom of the potential wells.

%%%%%%%%%%%%%%%%%%%%%%%%%%%%%%%%%%%%%%%%%%%%%%%
%%%%%%%%%%%%%%%%%%%%%%%%%%%%%%%%%%%%%%%%%%%%%%%

In our experiment $^{85}$Rb atoms are cooled and trapped in a magneto optical
trap (MOT). This is obtained by applying an inhomogeneous magnetic field, and
three orthogonal pairs of counterpropagating $\sigma^{\pm}$ laser fields. We
indicate by $x,y,z$ the propagation directions of these fields. At a given
instant the MOT magnetic field is turned off and the circularly polarized
laser fields along the $z$-axis are replaced by the two crossed polarized
lattice beams. The $\sigma^{\pm}$ laser fields in the $x$ and $y$ directions
are left on, so to provide a friction force in the direction orthogonal to the
one of the periodic potential. In this way the motion of the atoms in the $x$
and $y$ direction is damped, and the atomic dynamics in the $z$ direction
can be studied for longer times.  The appropriate (modulated) phase relation
between the two lattice fields (Eq. \ref{alpha}) is obtained by using two
acousto-optical modulators (AOM), one for each lattice beam (Fig. \ref{fig1}).
The AOMs are driven by radio-frequency generators oscillating at
$\Omega=76$ MHz and sharing the same reference clock. One of this
radio-frequency generator is phase-modulated by a signal obtained by mixing
the output of two oscillators at frequencies $\omega$ and $2\omega$
($\omega\simeq 100$ kHz) and phase difference $\phi$. These two oscillators
share the same reference clock.

\begin{figure}[ht]
\begin{center}
\mbox{\epsfxsize 3.in \epsfbox{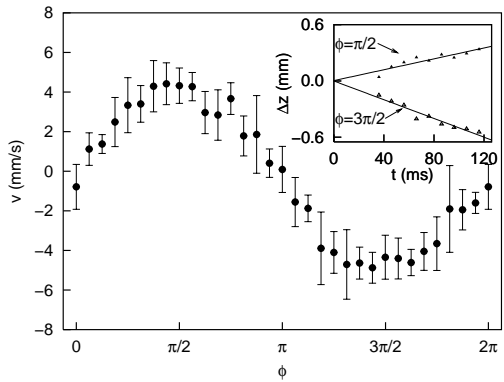}}
\end{center}
\caption{Velocity of the center-of-mass of the atomic cloud as a function of
the phase $\phi$. Inset: Displacement along the $z$ axis of the CM of the
atomic cloud as a function of the lattice duration for the two values of the
phase $\phi$ corresponding to the maximum velocity in the two opposite
directions ($\pm z$), together with the linear fits. The detuning of the
lattice fields from atomic resonance is $\Delta = 36$ MHz, the intensity per
lattice beam is $I_L=7$ mW/cm$^2$. For these parameters the oscillation
frequency of the atoms at the bottom of the potential well is $\Omega_v\simeq
105$ kHz.  The parameters for the phase-modulation signal $\alpha (t)$ (see
Eq. \protect\ref{alpha}) are $\omega=113$ KHz,
$A=3/4$, $B=1$ with $\alpha_0=10$ rad.}
\label{fig2}
\end{figure}

We studied the dynamics of the atoms in the optical lattice by direct
imaging with a Charge Coupled Device (CCD) camera. For a given phase $\phi$
we took images of the atomic cloud at different instants after the atoms
have been loaded into the optical lattice. From the images of the atomic
cloud we determined the position along the $z$ axis of the center-of-mass
(CM) of the atomic cloud as a function of the lattice duration. It should be
noted that for the typical time scales of our experiments the measured
position of the CM of the atomic cloud in the laboratory and in the accelerated
reference frames are approximately equal. In fact the accelerated frame
oscillates with an amplitude of about 1 $\mu$m, while the typical displacement
of the CM associated to the directed diffusion is 100 $\mu$m. Furthermore,
for a typical frequency $\omega\simeq 100$ kHz the position $z$ in the
laboratory frame and the corresponding position in the accelerated frame
$z'=z-\alpha(t)/(2k)$, with $\alpha (t)$ given by (\ref{alpha}), are
equivalent when averaged over a typical exposure time of $1$ ms. Therefore
for the measurement of the position of the CM of the atomic cloud
no coordinate transformation is needed to go from the laboratory frame to the
accelerated frame where the description in terms of a static potential and an
applied force is valid. We made several measurements for different values of
the phase $\phi$. We observed that the CM of the atomic cloud moves along the
$z$ axis with constant velocity, as shown in the inset of Fig. \ref{fig2}. We
determined the CM-velocity as a function of the phase
$\phi$, with results as in Fig. \ref{fig2}. The experimental results of
figures \ref{fig2} clearly demonstrate the phenomenon of
directed diffusion in a symmetric periodic potential: the atoms can be set
into a directed motion through a symmetric potential by breaking the temporal
symmetry of the system.

%%%%%%%%%%%%%%%%%%%%%%%%%%%%%%%%%%%%%%%%%%%%%%%
%%%%%%%%%%%%%%%%%%%%%%%%%%%%%%%%%%%%%%%%%%%%%%%

The dependence of the CM-velocity on the phase $\phi$, shown in
Fig. \ref{fig2}, can be explained by examining the temporal symmetries of the
system \cite{flach,flach2}. In fact although the symmetry $F(t+T/2) = - F(t)$
is broken for any value of the phase $\phi$,
there is an additional temporal symmetry $F(t)=F(-t)$, which implies zero
net current through the potential for particular values of $\phi$
\cite{flach,flach2}. This symmetry is
realized for $\phi=n\pi$, with $n$ integer, and maximally broken for
$\phi=(n+1/2)\pi$. This explains the observed dependence of the CM-velocity
on the phase $\phi$, and shows that in our system $\phi$ is the control
parameter of the directed diffusion.

%%%%%%%%%%%%%%%%%%%%%%%%%%%%%%%%%%%%%%%%%%%%%%%
\begin{figure}[ht]
\begin{center}
\mbox{\epsfxsize 3.in \epsfbox{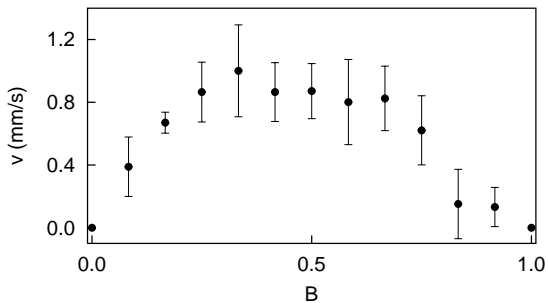}}
\end{center}
\caption{Velocity of the center-of-mass of the atomic cloud as a function of
the amplitude $B$ of the component at $2\omega$ of the driving force, for
constant sum of the amplitudes of the two harmonics at $\omega$ and $2\omega$.
The parameters for the optical lattice are the same as for Fig.
\protect\ref{fig2}. The phase-modulation signal is given by Eq.
(\protect\ref{alpha}) with $A=1-B$, $\omega=100$ KHz, $\alpha_0=12$ rad
and $\phi=\pi/2$.}
\label{fig3}
\end{figure}
%%%%%%%%%%%%%%%%%%%%%%%%%%%%%%%%%%%%%%%%%%%%%%%

To demonstrate experimentally that directed diffusion is determined by the
breaking of the symmetry $F(t+T/2)=-F(t)$, we fix the phase $\phi$ equal to
$\pi/2$, so to maximally break the $F(t)=F(-t)$ symmetry, and study the 
CM-velocity as a function of the amplitudes of
the harmonics of frequencies $\omega$ and $2\omega$ of the driving force. 
We choose a phase-modulation of the form of Eq. (\ref{alpha}) with
$A=1-B$:
$\alpha (t)=\alpha_0[(1-B)\cos (\omega t) +B/4\cos (2\omega t-\phi)]$,
so to obtain a force 
$F=M\omega^2\alpha_0/2k [(1-B)\cos (\omega t)+B \cos (2\omega t-\phi )]$.
Thus, by varying the parameter $B$ we vary the ratio of the amplitudes
of the two components of the force at frequencies $\omega$ and $2\omega$,
while keeping constant their sum. The experimental results are shown in 
Fig. \ref{fig3}. We observe that for $B=0$ and $B=1$, which correspond
to a monochromatic driving force, there is no net transport of atoms. By 
increasing B from the zero value the atoms are set into directed
motion, and a maximum for the CM-velocity is reached for $B\simeq 0.5$,
i.e. for about equal amplitudes of the even and odd harmonics. This
demonstrates that DD is determined by the breaking of the symmetry 
$F(t+T/2)=-F(t)$.

%%%%%%%%%%%%%%%%%%%%%%%%%%%%%%%%%%%%%%%%%%%%%%%
\begin{figure}[ht]
\begin{center}
\mbox{\epsfxsize 3.in \epsfbox{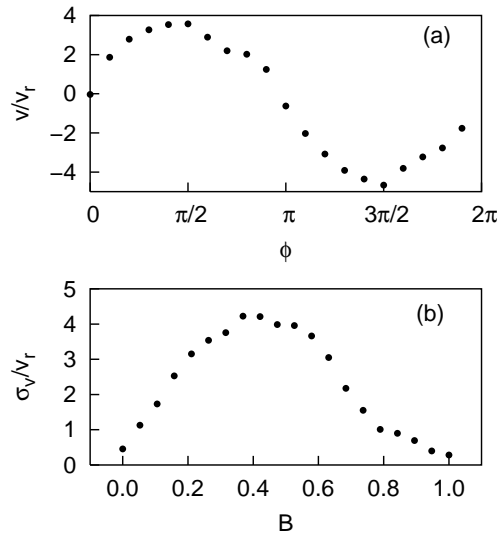}}
\end{center}
\caption{Results of semiclassical Monte Carlo simulations for the atomic
dynamics in the 1D-lin$\perp$lin optical lattice. The phase-modulation
$\alpha (t)$ has the form of Eq. (\protect\ref{alpha}), with $A=1-B$,
$\omega=0.87\Omega_v$.
The lattice
parameters are: light shift per beam $\Delta_0'=-150\omega_r$ and lattice
detuning $\Delta = -5 \Gamma$. Here $\Gamma$ and $\omega_r$ are the width
of the excited state and the atomic recoil frequency, respectively.
In (a) the CM-velocity in units of the atomic recoil velocity ($v_r$),
is plotted as a function of the phase
$\phi$, for $\alpha_0=8$ and $B=1/2$.
In (b) the amplitude $\sigma_v$ of the velocity-curve is plotted
as a function of the amplitude $B$ of the component at $2\omega$ of the
driving force, for constant sum of the amplitudes of the two harmonics at
$\omega$ and $2\omega$. Here $\alpha_0=3$.}
\label{fig4}
\end{figure}
%%%%%%%%%%%%%%%%%%%%%%%%%%%%%%%%%%%%%%%%%%%%%%%

The microscopic mechanism producing a nonzero current of atoms through
the optical lattice can be related to the general mechanism of current
rectification following harmonic mixing firstly evoked to explain the
electronic transport properties of crystals \cite{seeger}, and recently 
reexamined (Ref. \cite{flach2} and references therein). For the specific
system considered in the present work, the harmonic mixing results in a
displacement $\Delta z$ of the center of oscillation $\langle z(t)\rangle$
of the atoms in a potential well from the well center. Such a 
displacement originates from the anharmonicity of the potential, and it
is quadratic in the amplitude of the field at frequency $\omega$ and 
linear in the amplitude at $2\omega$: $\Delta z\propto A^2 B$. Therefore
a nonzero $\Delta z$ is obtained only when both components of the force
are applied.  As the optical pumping
rate $\Gamma'$ (escape rate) toward neighbouring wells increases with 
the distance from the well center ($\Gamma' \propto \sin^2 k \Delta z$,
see Ref. \cite{robi}), such a displacement results in an asymmetry between
the escape rates toward the left and right wells, and a nonzero current
of atoms is produced.

Our experimental observations are supported by semiclassical Monte Carlo
simulations for a $J_g=1/2\to J_e=3/2$ atomic transition.  We examined
the atomic dynamics in the 1D-lin$\perp$lin optical lattice for a phase
modulation of one of the lattice beams of the form (\ref{alpha}).  For
given amplitudes of the even and odd harmonics, we calculate the CM-velocity
as a function of the phase $\phi$, with results as in Fig. \ref{fig4}(a).
The data are in complete agreement with the experimental findings, and 
confirm that the cloud of atoms is set into directed motion whenever the
temporal symmetry $F(t)=F(-t)$ is broken.

The amplitude of the velocity-curves as the one of Fig. \ref{fig4}(a) has
been characterized by the quantity
$ \sigma_v=[( \sum_{i=1,N} v^2_{\phi_i}-\langle v\rangle )/N]^{1/2}$
where $v_{\phi_i}$, with $i=1,..., N$ are the numerical results for the
CM-velocity at the phase $\phi=\phi_i$. By plotting (see Fig. \ref{fig4}(b))
the quantity $\sigma_v$ as a function of the amplitude $B$ of the component
at $2\omega$ of the driving force, for constant sum of the amplitudes of
the two harmonics at $\omega$ and $2\omega$ we recover the behaviour
observed in the experiment: a non-zero value of the amplitude $B$ corresponds
to the breaking of the $F(t+T/2)=-F(t)$ symmetry, and leads to the directed
motion of the atoms.

In conclusion, in this work we demonstrated experimentally the phenomenon of
directed diffusion in a symmetric periodic potential. This has been
demonstrated with cold atoms in a periodic optical lattice.
The same sort of behaviour was previously obtained in an asymmetric
periodic potential (ratchet) \cite{robi99}. The symmetric periodic potential
corresponds to a 1D-lin$\perp$lin optical lattice. Two counterpropagating
laser fields produce both the periodic potential and a friction force for the
atoms. Furthermore the stochastic process of optical pumping leads to a
diffusive dynamics of the atoms through the periodic structure. A force of
zero average is applied by phase-modulating one of the lattice fields. Indeed,
in an accelerated frame the atoms see a static symmetric periodic potential
and an inertial force which breaks the temporal symmetry of the system. The
degree of temporal symmetry-breaking of the system can be carefully controlled
by varying the parameters of the phase modulation determining the force in the
noninertial reference frame.  We demonstrated that the atoms can be set into
directed motion by breaking the temporal symmetry of the system.

The present realization of directed diffusion has been obtained in the 
regime of {\it non-adiabatic} driving, i.e. for a driving force of about
the same frequency of the oscillations of the atoms at the bottom of the
potential wells. This qualifies our system as testing ground for the recent
theory of resonant activation based on logarithmic susceptibilities
\cite{rabitz,rabitz2}.

Laboratoire Kastler Brossel is an "unit\'e mixte de recherche de l'Ecole
Normale Sup\'erieure et de l'Universit\'e Pierre et Marie Curie associ\'ee
au Centre National de la Recherche Scientifique (CNRS)".

\end{document}